\documentstyle[aps,pra,epsf,multicol]{revtex}
\begin{document}
\title{Communication Capacity of Quantum Computation}
\author{S. Bose, L. Rallan and V. Vedral}
\address{Centre for Quantum Computation, Clarendon Laboratory,
	University of Oxford,
	Parks Road,
	Oxford OX1 3PU, England}
\date{\today}
\maketitle
\begin{abstract}
By considering quantum computation as a communication process, we relate 
its efficiency to a communication capacity.
This formalism allows us to rederive lower bounds on the 
complexity of search algorithms. It also enables us to link the
mixedness of a quantum computer to its efficiency. We discuss
the implications of our results for quantum measurement.
\end{abstract}
\draft
\begin{multicols}{2}
Any computation, both classical and
quantum, is formally identical to a communication in time.
At time $t=0$, the programmer sets the computer to
accomplish any one of several possible
tasks. Each of these tasks can be regarded as embodying
a different message. Another programmer can obtain this
message by looking at the output of
the computer when the computation is finished at time $t=t_1$.
Recent years have witnessed a surge of interest in
both quantum computation \cite{Deu85,Shor,Grov} and quantum
communication \cite{khol,BB84}. 
Computation based on quantum principles allows for more efficient
algorithms for solving certain problems than
algorithms based on purely classical principles. Quantum communication,
on the other hand, can be used for unconditionally secure secret
key distribution \cite{Deu96}. However, till date, these two
areas (i.e quantum computation and quantum communication) have
developed independently. In this letter we connect the classical
capacity of a quantum communication channel \cite{khol} with the efficiency
of quantum computation. This offers an unifying framework
for quantum information processing.

  Let us first introduce a few definitions and a communication model of
quantum computation.  We have two programmers, the sender
and the reciever and two registers, the memory ($M$) register and
the computational ($C$) register.  The sender prepares the memory register in a certain
quantum state $|i\rangle_{\scriptsize M}$ which encodes the problem to be solved. For example, in the
case of factorization \cite{Shor}, this register will store the number
to be factored. In case of a database search \cite{Grov}, this register will
store the state of the database to be searched. The number $N$ of possible states $|i\rangle_{\scriptsize M}$ will,
of course, be limited by the greatest number that the given computer
could factor or the largest database it could search. The reciever
 prepares the computational register in some initial state $\rho^0_{\scriptsize C}$. Both the sender and the reciever feed the registers (prepared by
them) to the quantum computer. The
quantum computer implements the following general transformation
on the registers
\begin{equation}
 (|i\rangle \langle i|)_{\scriptsize M} \otimes \rho^0_{\scriptsize C} \rightarrow (|i\rangle \langle i|)_{\scriptsize M} \otimes U_i \rho_{\scriptsize C} U_i^{\dagger}.
\end{equation}
The resulting state  $\rho_{\scriptsize C}(i)
= U_i \rho^0_{\scriptsize C} U_i^{\dagger}$ of the computational register
contains the answer to the computation and is measured by the reciever.
As the quantum computation should work for any $|i\rangle_M$, it should
also work for any mixture $\sum_i^N p_i (|i\rangle\langle i|)_{\scriptsize M}$,
where $p_i$ are probabilities.
For the sender to use the above
computation as a communication protocol, he has to prepare
any one of the states $|i\rangle_{\scriptsize M}$ with an {\em apriori} probability
$p_i$. The entire input ensemble is thus $\sum_i^N p_i (|i\rangle\langle i|)_{\scriptsize M}
\otimes \rho^0_{\scriptsize C}$. Due to the quantum computation, this becomes

\begin{equation}
 \sum_i^N p_i (|i\rangle\langle i|)_{\scriptsize M}
\otimes \rho^0_{\scriptsize C} \rightarrow \sum_i^N p_i (|i\rangle\langle i|)_{\scriptsize M}
\otimes \rho_{\scriptsize C}(i).
\end{equation}
Whereas before the quantum computation,
the two registers where completely uncorrelated (mutual information
is zero), at the end, the
mutual information becomes
\begin{eqnarray}
\label{mut}
I_{MC}:&=& S(\rho_{\scriptsize M}) + S(\rho_{\scriptsize C}) -S (\rho_{\scriptsize MC}) \nonumber \\
   &=& S(\rho_{\scriptsize C}) - \sum_i^N p_i S(\rho_{\scriptsize C}(i)),
\end{eqnarray}
where $\rho_{\scriptsize M}$ and $\rho_{\scriptsize C}$ are the reduced density operators for the
two registers, $\rho_{\scriptsize MC}$ is the density operator of entire $M+C$ system and
$S(\rho)=-\mbox{Tr}\rho \log \rho$ is the von Neumann entropy (for conventional
reasons we will use $\log_2$ in all calculations). Notice that
the value of the mutual information (i.e correlations) is equal to
the Holevo bound $H=S(\rho_{\scriptsize C}) - \sum_i^N p_i S(\rho_{\scriptsize C}(i))$ for the
classical capacity of a quantum communication channel \cite{khol}
(Note that $\rho_{\scriptsize C}=\sum_i^N p_i \rho_{\scriptsize C}(i)$). This tells us how
much information the reciever can obtain about the choice $|i\rangle_M$ 
made by the sender by measuring the computational register. The maximum
value of $H$ is obtained when the states $\rho_{\scriptsize C}(i)$ are pure and orthogonal.
Moreover, the sender conveys the maximum information when all the
message states have equal {\em apriori} probability (which also maximizes the
channel capacity). In that case the mutual information (channel capacity)
at the
end of the computation is
$\log{N}$. Thus the communication capacity $I_{MC}$ (given by Eq.(\ref{mut}))
gives an index of the efficiency of a quantum computation.
{\em The target of a quantum computation is to achieve the maximum
possible communication capacity consistent with given initial
states of the quantum computer}. If one breaks down the general unitary transformation $U_i$ of a quantum algorithm into several 
succesive unitary transformations, then the maximum
capacity may be achieved only after several steps. In each of the smaller unitary transformations, the mutual information between the $M$ and the
$C$ registers (i.e the communication capacity) increases by a certain amount. When its total value
reaches the maximum possible value consistent with a given initial state
of the quantum computer, the computation is regarded as being complete. 

  We now proceed to illustrate one immediate application of the above
formalism. Any general quantum algorithm has to have a certain
number of queries into the memory register \cite{Bennett,Beals,Ambainis}
(this is neccessiated by the fact that the transformation on the computational
register has to depend on the problem at hand, encoded in $|i\rangle_{\scriptsize M}$). These queries can be considered to
be implemented
by a black box into which the states of both the memory and the
computational registers are fed. 
The number of such queries needed
in a certain quantum algorithm gives the black box complexity of
that algorithm \cite{Bennett,Beals,Ambainis} and is a
lower bound on the complexity of the whole
algorithm. Recently, Ambainis \cite{Ambainis}
showed in a very elegant paper that if the memory register was prepared initially in the superposition
$\sum_i^N |i\rangle_{\scriptsize M}$, then, in a search
algorithm, $O(\sqrt{N})$ queries
would be needed to completely entangle it with the 
computational register. This gives a lower bound
on the number of queries in a search algorithm.
In a manner analogous to his, we will calculate the change
in mutual information between the memory and the computational
registers (from Eq.(\ref{mut})) in one query step. The number
of queries needed to increase the mutual information to $\log{N}$
(for perfect communication between the sender and the reciever), is then
a lower bound on the complexity of the algorithm.

    Any search algorithm (whether quantum or classical, irrespective
of its explicit form), will have to find a match for the state $|i\rangle_{\scriptsize M}$ of the $M$ register among the states
$|j\rangle_C$ of the $C$ register and associate a marker
to the state that matches (Here, $|j\rangle_C$ is a complete
orthonormal basis for the $C$ register). The most general way of doing
such a query in the quantum case 
is the black box unitary transformation \cite{Ambainis}  
\begin{equation}
 U_{\scriptsize B}|i\rangle_M |j\rangle_C = (-1)^{\delta_{ij}} |i\rangle_M |j\rangle_C.
\end{equation}
Any other unitary transformation performing a query matching
the states of the $M$ and the $C$ registers, could be constructed from the above
type of query. 
We would like to put a bound on the change of the mutual information in
one such black box step. Let the memory states $|i\rangle_M$ be
available to the sender with equal apriori probability so that the
communication capacity is a maximum. His initial ensemble
is then $\frac{1}{N}\sum_{i}^N (|i\rangle \langle i|)_{\scriptsize M}$.
Let the reciever prepare the $C$ register in an initial
pure state $\psi^0$ (in fact, the power of quantum computation stems
from the ability of the reciever to prepare pure state
superpositions of form $\frac{1}{N}\sum_{j}^N |j\rangle_C$).
In general, there will be many black box steps on the
initial ensemble before perfect correlations between the $M$ and the $C$
registers is set up.
 Let, after the $k$th black box step, the state of the
system be
\begin{equation}
\rho^k = \frac{1}{N}\sum_{i}^N (|i\rangle \langle i|)_{\scriptsize M} \otimes (|\psi^k(i)\rangle \langle \psi^k(i)|)_{\scriptsize C}
\end{equation}
where
\begin{equation}
|\psi^k(i)\rangle_{\scriptsize C} = \sum_j \alpha_{ij}^k |j\rangle_{\scriptsize C}.
\end{equation}
The $(k+1)$th black box step changes this state to $\rho^{k+1}=\frac{1}{N}\sum_{i}^N (|i\rangle \langle i|)_{\scriptsize M} \otimes (|\psi^{k+1}(i)\rangle \langle \psi^{k+1}(i)|)_{\scriptsize C}$ with
  \begin{equation}
|\psi^{(k+1)}(i)\rangle = \sum_{i,j}^{N} \alpha_{ij}^k (-1)^{\delta_{ij}}|i\rangle_{\scriptsize M} |j\rangle_{\scriptsize C}.
\end{equation}
 Thus we
only have to evaluate the difference of mutual information between
the $M$ and the $C$ register for the states.  
This difference of mutual information (when computed from Eq.(\ref{mut}))
can be shown to be the difference $|S(\rho^{k+1}_{\scriptsize C})-S(\rho^k_{\scriptsize C})|$ \cite{Hend}. This quantity is bounded from
the above by \cite{Fannes}
\begin{eqnarray}
 |S(\rho^{k+1}_{\scriptsize C}) &-& S(\rho^k_{\scriptsize C})| \leq
d_{\scriptsize B}(\rho^k_{\scriptsize C},\rho^{k+1}_{\scriptsize C}) \log{N} \nonumber \\ &-& d_{\scriptsize B}(\rho^k_{\scriptsize C},\rho^{k+1}_{\scriptsize C}) \log{d_{\scriptsize B}(\rho^k_{\scriptsize C},\rho^{k+1}_{\scriptsize C})}
\end{eqnarray}
where, $d_{\scriptsize B}(\sigma,\rho)=\sqrt{1-F^2(\sigma,\rho)}$ is the
Bures metric and $F(\sigma,\rho)=\mbox{Tr}\sqrt{\sqrt{\rho}\sigma \sqrt{\rho}}$
is the fidelity. Using methods similar to Ambainis \cite{Ambainis}, it
can be shown that $F(\rho^k_{\scriptsize C},\rho^{k+1}_{\scriptsize C}) \geq \frac{N-2}{N}$ from which it follows that
\begin{equation}
\label{step}
 |S(\rho^{k+1}_{\scriptsize C})-S(\rho^k_{\scriptsize C})| \leq \frac{3}{\sqrt{N}} \log{N}.
\end{equation}
 This means that at least $O(\sqrt{N})$ steps are needed to produce
full correlations (maximum mutual information of value $\log{N}$)
between the two registers. This gives the black box lower
bound on the complexity of
any quantum search algorithm. Of course, we know that there also exists
an algorithm achieving this bound due to Grover \cite{Grov} and this has
been proven to be optimal \cite{Bennett,Ambainis,zalka}.

\begin{figure}
\begin{center} 
\leavevmode 
\epsfxsize=8cm 
\epsfbox{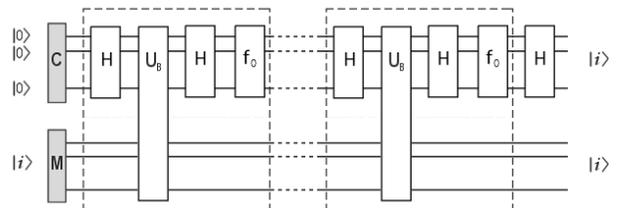}
\caption{\narrowtext The figure shows the circuit for Grover's algorithm.
$C$ is the computational register and $M$ is the memory register. $U_{\scriptsize B}$ is the
black box query transformation, $H$ is a Hadamard transformation
on every qubit of the $C$ register and $f_0$ is a phase flip in front
of the $|00...0\rangle_{\scriptsize C}$. The block consisting of $H,U_{\scriptsize B},H$ and $f_0$ is repeated a number of times.}
\label{setup} 
\end{center}
\end{figure}

\begin{figure}
\begin{center} 
\leavevmode 
\epsfxsize=8cm 
\epsfbox{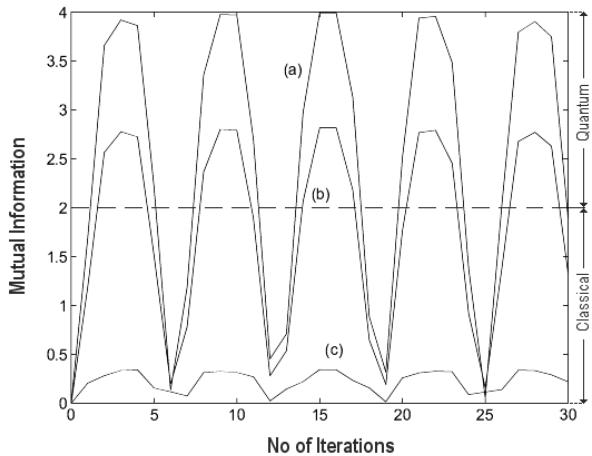}
\caption{\narrowtext The figure shows the dependence of the mutual information
between the $M$ and the $C$ registers as a function of the number of times
the block in Grover's algorithm is iterated for various values of initial
mixedness of the $C$ register. Each qubit of the $C$ register is initially 
in the state $p|0\rangle\langle 0|+(1-p)|1\rangle \langle 1|$, 
(a) $p=1$, (b) $p=0.95$
and (c) $p=0.7$. The (a) and (b) computations achieve higher
mutual information  than classically
allowed in the order of root $N$ steps, while (c) does not.}
\label{graph} 
\end{center}
\end{figure}

   We now use Grover's algorithm to show how the mutual information varies
with time in a quantum search. 
The general sequence described by Cleve et. al \cite{mosca}
for Grover's algorithm will be used in this letter.  The algorithm
consists of repeated blocks, each consisting of a Hadamard transform on each
qubit of the $C$ register, followed by a $U_{\scriptsize B}$ (our
black box transformation), followed by another Hadamard transform on each
qubit of the $C$ register and finally a phase flip $f_0$ of the
the $|00...0\rangle_{\scriptsize C}$ state of the $C$
register (See fig.\ref{setup}). 
This block can then be repeated as many times as is necessary
to bring the mutual information to its maximum value of
$\log{N}$, which, as we have shown in Eq.(\ref{step}) to be $O(\sqrt{N})$.
Note that the only transformation correlating the $M$ and $C$
registers is the black box transformation $U_{\scriptsize B}$ and all
the other transformations are done {\em only} on the $C$ register and
therefore do not change the mutual information between the two registers.
In fig.\ref{graph} we have plotted the variation of mutual information
between the $M$ and the $C$ registers (i.e the communication capacity of
the quantum computation)
with the number of iterations of the block in Grover's algorithm. It is
seen that the mutual information oscillates with the number of iterations.
Fig.\ref{graph} is plotted for a four qubit computational register which
can search a database of $16$ entries. It is seen that the period is
roughly $6$, which means that the number of steps needed to achieve
maximum mutual information is roughly $3$. This is well above our bound
for the minimum number of steps, which is $4/3$ in this case.

  The three graphs (a), (b) and (c) in Fig.\ref{graph} are for different values
of initial mixedness of the $C$ register. 
We find that the mutual information fails to rise to the maximum
value of $\log{N}$ when the state of the computational register is
mixed. Our formalism thus allows us
to calculate the performance of a quantum computation as a function of the
mixedness (quantified by the von Neumann entropy) of the computational
register. We can put a bound on the entropy of the second register
after which the quantum search becomes as inefficient as the classical search.
If the initial entropy $S(\rho^{0}_{\scriptsize C})$ of the $C$ register exceeds $\frac{1}{2}\log{N}$, then the change in mutual information between
the $M$ and the $C$ registers in the course of the entire
quantum computation would be at most $\log{\sqrt{N}}$. This can be
achieved by a classical database search in $\sqrt{N}$ steps. So there is
no advantage in using quantum evolution when the initial state is too
mixed. Note that our condition 
\begin{equation}
S(\rho^{0}_{\scriptsize C}) \geq \frac{1}{2}\log{N}
\end{equation}
for no quantum speedup in the search
algorithm is only a sufficient condition and not a neccessary condition. This is similar to the
entropic conditions sufficient to ensure no quantum benefit from
teleportation and dense coding \cite{mix}. Analogous analysis can be
applied to any other algorithm.

   Finally, we point out that the states of the $M$ register
need not be a mixture, but could be
an arbitrary superposition of states $|i\rangle_{\scriptsize M}$
(such a state was used by Ambainis in his argument \cite{Ambainis}).
All the above arguments still hold in that case, and the $M$ and
the $C$ registers become quantum mechanically
entangled and not just classically correlated.
Thus our analysis implies that 
any quantum
computation is mathematically identical to
a measurement process \cite{everett}. The system being measured is the 
$M$ register and the apparatus is the $C$ register of the quantum
computer. As the time progresses the apparatus (register $C$) becomes more and more
correlated (or entangled) to the system (register $M$). This means that the
 states of register $C$ become more and more distinguishable which allows us
to extract more information about the $M$ register by measuring the $C$
register. The analysis in the last paragraph, where we showed the limitations
on the efficiency of quantum computation imposed by the mixedness of the
$C$ register, applies also to the efficiency of a quantum measurement when
the apparatus is in a mixed state. Mixedness of an apparatus, to the best of our
knowledge,
has never been considered in the analysis of quantum measurement. In general
practice, any
apparatus, however macroscopic, is considered to be in a pure quantum state
before the measurement. Our approach highlighting the formal analogy
between measurement and computation offers a way to analyse measurement
in a much more general context.

 L.R. would like to thank Invensys Plc for financial support.

     \end{multicols}
\end{document}